# Experimental observation of robust surface states on photonic crystals possessing single and double Weyl points

Wen-Jie Chen, Meng Xiao and C. T. Chan

Department of Physics and Institute for Advanced Study, Hong Kong University of Science and Technology, Hong Kong, China

* Email: phchan@ust.hk

We designed and fabricated a time-reversal invariant Weyl photonic crystal that possesses single Weyl nodes with topological charge of 1 and double Weyl nodes with a higher topological charge of 2. Using numerical simulations and microwave experiment, nontrivial band gaps with nonzero Chern numbers for a fixed $k_z$ was demonstrated. The robustness of the surface state between the Weyl photonic crystal and PEC against $k_z$-conserving scattering was experimentally observed.

PACS numbers: 41.20.Jb, 42.70.Qs

Stimulated by the concepts of quantum Hall effect [1,2] and topological insulators [3,4], the analogues of two-dimensional (2D) quantum Hall effect and quantum spin Hall effect have been proposed and realized in various kinds of systems [5-31]. These systems have nontrivial bulk band gap characterized by nonzero (spin) Chern number and gapless edge states on their boundaries. Recently, researchers began to explore topological states in three-dimensional classical wave systems, such as Weyl photonic crystals [32,33] and photonic topological crystalline insulators [34]. In contrast to topological insulator, Weyl photonic crystal, as its electronic counterpart [35-37], has a gapless bulk band structure with pairs of band degeneracy points (namely Weyl points) and topologically-protected surface states. In the neighborhood of these Weyl points with charge of ±1, two bands touch linearly in all the three directions, which were observed very recently by angle-resolved spectrum measurement in both electronic [38-39] and photonic [33] system. These Weyl points carry topological charges and serve as quantized sources or sinks of Berry flux in 3D reciprocal space, which predicts the existence of surface states connecting Weyl points with opposite charges. Recently, double Weyl points with topological charges of 2 are also discovered [40]. In addition, the concept of Weyl point has been extended to acoustic system [41] by engineering interlayer coupling. In their theoretical proposal, the chiral interlayer coupling mimics an effective gauge flux when the system is reduced to 2D with fixed $k_z$.

In this paper, we designed, fabricated and experimentally characterized a Weyl photonic crystal by introducing chiral interlayer coupling in microwave regime. The system carries both single and double Weyl points. Nontrivial 2D bulk band gaps for fixed values of $k_z$ and Weyl points were confirmed by angle-resolved transmission spectra. The robustness of the associated surface states against $k_z$-preserved scattering were experimentally observed.

It is known that Weyl points in acoustic meta-crystals [41] can be realized theoretically using multilayers of honeycomb lattices stacked in the z-direction, with two of the Weyl points locate at K and K'. The acoustic meta-crystal was designed [41] using a nearest-neighbor tight-binding model in which the in-plane honeycomb lattice gives rise to the conical dispersion near Brillouin zone corners (K and K') in the $k_x$-$k_y$ plane, while the chiral interlayer hopping gives rise to the linear dispersion in the $k_z$ direction. Here, the tight-binding Hamiltonian is not used as a design tool per se, but rather as a starting point to guide us to the structures that have the correct symmetry to support topological features such as synthetic gauge flux and associated Weyl points. This can facilitate the design and fabrication of our microwave sample. Figure 1(a) depicts the unit cell of a single layer system, which is a hexagonal array of perfect electric conductor (PEC) cylinders embedded in a parallel plate (PEC) waveguide. Its fundamental waveguide mode is $E_z$-polarized and has a conical dispersion at K and K'. This planar waveguide can be achieved by a printed circuit board (PCB), with both sides cladded by copper, pierced by a hexagonal array of metal cylinders. One can stack the PCBs in the z-direction to form a 3D photonic crystal, as shown in Fig. 1(b). The PCBs are $0.43a$-thick and their spacing in the z-direction is $0.11a$, where $a = 1cm$ is the lattice constant of the hexagonal array. The dielectric constant of FR4 substrate (yellow in Fig. 1(b)) is 4.8 and the thickness of deposited copper layer (shown in gray and assumed to be PEC in our simulation) is $40\mu m$. This structure has a 3D hexagonal lattice with in-plane lattice constant $a$ and out-of-plane lattice constant $d = 0.54a$. Figure 1(c) shows its 3D Brillouin zone. Interlayer coupling are introduced by etched Y-shaped slots on the top and bottom copper layers. Figure 1(d) shows the top view of the unit cell (dashed hexagon), where the Y slots on top and bottom surfaces are shown in blue and red respectively. It can be seen that the slots' patterns are chiral and break all the mirror symmetries and inversion symmetry.

The bulk band structure at the $k_z = 0$ plane and the one from k-points –H to H are shown in Fig. 1(e) and Fig. 1(i), respectively. We note that the metallic components (rods and copper layers) in the photonic crystal form a connected metallic network, which leads to a cutoff frequency of 9.23 GHz for the bulk modes. One finds that the first and the second, the fifth and the sixth bands touch at K point, and the dispersions near the degeneracy point are linear in all three directions of the reciprocal space, indicating that they are Weyl points. The topological charge of Weyl point can be calculated by either integrating Berry curvature on a closed surface enclosing the Weyl point or inspecting the rotational eigenvalues of the two touching bands [40]. The topological charges for the Weyl points at 12.25 and 9.94 GHz at K are -1 and 1, which are labeled by the colored circles in Fig. 1(e). Due to the $C_6$ symmetry of the photonic crystal, Weyl points at K' point with same charges are expected. In addition, we also find another three degeneracy points at $\Gamma$ (highlighted by colored circles). From Fig. 1(e) and 1(g), we see that the band dispersions near these points are linear along $k_z$-direction and quadratic in $k_x$-$k_y$ plane. They are double Weyl points [40] with charge of ±2 which are the superposition of two single Weyl points with the same charge of ±1. To the best of our knowledge, this is the first time that a real structure has been designed and fabricated to exhibit double Weyl points with topological charges of ±2. It has been predicted that double Weyl points can exist in crystals possessing $C_4$ or $C_6$ point group symmetries [40]. In our case, the degeneracy between the two single Weyl points is protected by $C_3$ symmetry. Once the $C_3$ symmetry is broken, two Weyl points with charge of ±1 will separate and each forms a linear dispersion in all three directions. Here we give an example of $C_3$ symmetry-broken photonic crystal. Figure 2(a) shows the top view of the unit cell, which is the same as in Fig. 1(d) except that the circular cylinder with radius of $0.2a$ is replaced by an elliptical cylinder. Figures 2(b) and 2(c) plot the corresponding Brillouin zone and band structure at the $k_z = 0$ plane. Each of the three quadratic dispersive double Weyl points in Fig. 1(e) at Γ breaks into a pair of single Weyl points of the same topological charge. Three of these single Weyl points show up as linear dispersive crossing points at $k_x$ (along Γ–M direction) or $k_y$ (along Γ–K direction) axis in Fig. 2c (the degeneracy points between second/third bands and fourth/fifth bands along Γ-M, and between sixth/seventh bands along Γ-K). These are single Weyl points have linear dispersion in all three directions. For example, Figures 2(d) and 2(e) show the dispersions near the crossing point at $(0.35\pi/a, 0, 0)$, in $k_y$ and $k_z$ directions, respectively. It can be inferred that another single Weyl point with the same topological charge should lie at $(-0.35\pi/a, 0, 0)$ by applying $C_2$ rotation. In addition, the two single Weyl points, originally at K', shift toward Γ point due to the $C_3$ symmetry-breaking. This manifests the robustness of Weyl points. They cannot be gapped easily and a perturbation just shift their positions in k-space. To verify that the double Weyl is protected by $C_3$ symmetry, we also consider a photonic crystal with the circular rods replaced by triangular rods which reduces the symmetry from $C_6$ to $C_3$, and the double Weyl points are founded in the band structure (see the Supplemental Material).

Since the crystal has translational symmetry along z-direction, $k_z$ is a good quantum number as long as the symmetry is preserved. If we fix a $k_z$ value and consider the dispersion and transport in x-y plane, the system is effectively reduced to 2D with a 2D band structure on a constant $k_z$ plane (the gray translucent plane in Fig. 1(c)). The $k_z$ is then a parameter that characterizes this 2D system. The Chern number of a nondegenerate 2D band for a fixed $k_z$ is well-defined. For example, Figure 1(f) calculates the bulk band structure when $k_z = 0.05\pi/d$. Compared with the band structure of $k_z = 0$, the band degeneracies at $\bar{\Gamma}$ and $\bar{K}$ are lifted. All the bands on $k_z = 0.05\pi/d$ are separated, and their Chern number can be calculated by analyzing the rotational eigenvalues at high symmetric k-points [42]. Specifically, we have $\exp(i2\pi C_n/6) = \eta_n(\bar{\Gamma})\theta_n(\bar{K})\zeta_n(\bar{M})$, where $C_n$, $\eta_n$, $\theta_n$ and $\zeta_n$ are the Chern number, $C_6$, $C_3$ and $C_2$ rotational eigenvalues of the nth band, respectively. Corresponding Chern numbers are labeled by gray numbers near the bands in Fig. 1(f). We found that a nontrivial band gap with a gap Chern number of 1 opens near 12.25 GHz, which will be observed in our experiment. By tuning the parameter $k_z$, the eigenmodes with different representations or rotational eigenvalues can exchange their positions, which leads to the change of band Chern numbers. Figure 1(g-i) plots the dispersion along the $k_z$ direction at $\bar{\Gamma}$, $\bar{M}$ and $\bar{K}$, respectively, where the bands with different rotational eigenvalues are plotted in different colors. It can be seen from Fig. 1(g) and 1(i) that several band crossings occur between two bands with different representations. Since these band inversions induce the changes of band Chern number or Berry curvature, the crossing points at the rotation axes should be Weyl points and carry Berry curvature's charge which is associated with the jump of Chern numbers [37]. For instance, the black dashed line in Figure 1(j) plots the Chern number of the fifth band as a function of parameter $k_z$. It presents several jumps, which coincide with the band inversions occurring at $\bar{\Gamma}$ and $\bar{K}$. The jumps at $k_z = 0$ and $k_z = \pi/d$ are due to the band inversion at $\bar{\Gamma}$ with the fourth band and the inversion at $\bar{K}$ with the sixth band. The jumps at $k_z = \pm 0.53\pi/d$ and $k_z = \pm 0.91\pi/d$ are due to the band inversions at $\bar{K}$ with the fourth band, which are labeled by colored circles in Fig. 1(i). The gap Chern number of the fifth gap, which is the summation of the band Chern numbers below the gap, is illustrated by red solid line in Fig. 1(j). It is +1 (-1) when kz is positive (negative), indicating the existence of the surface state propagating clockwise (anticlockwise) in x-y plane at the boundary of the crystal.

    To experimentally confirm the existence of Weyl points, we measured angle-resolved transmission of the Weyl photonic crystal. Two samples were fabricated, one with the surface perpendicular to the $\bar{\Gamma}-\bar{K}$ direction (as shown in Fig. 4(b)), the other perpendicular to the $\bar{\Gamma}-\bar{M}$ direction (Fig. 4(g)). Both samples consist of 60 PCBs stacked in the z-direction, which are supported by two plexiglass and pierced by aluminum rods. Electromagnetic (EM) waves are emitted from the left horn antenna with the electric field polarized in the x-z plane(see Fig. 3), and received by the right antenna with the same polarization. When an EM wave with angle $\theta$ and frequency $f$ impinges onto the sample in Fig. 4(b), all the bulk modes with $k_y = 0$ and $k_z = 2\pi f \sin\theta/c$ will be excited, as shown by the blue dashed line in Fig. 4(a). One can change $\theta$ by rotating the sample and scan the bulk modes with different $k_z$. Gray areas in Fig. 4(c) are the calculated bulk band projected along $\bar{\Gamma}-\bar{K}$ direction. The red line marks the $k_z$ corresponding to $\theta = 50°$, which is the maximal incident angle in our measurement. From Figs. 1(g) and 1(i), we see that the directional band gap between the fifth and the sixth bands in the $\bar{\Gamma}-\bar{K}$ direction opens at

nonzero $k_z$ and becomes larger as $k_z$ increases when $k_z < 0.3\pi/d$. This can also be seen in the projected band in the white region at about 12 GHz, together with the linear dispersion of the Weyl point at K. Figure 4(d) and 4(e) show the simulated and measured transmission spectra with different incident angle. Both agree well with the projected band. And the linear cone was observed which verifies the existence of Weyl point. In our COMSOL simulations, periodic boundary conditions are applied in the z-direction. The deviation of measured transmissions from simulated results may stem from the finite sizes of sample and incident beam.

In contrast to the $\overline{\Gamma}-\overline{K}$ direction, one finds from Figs. 1(g) and 1(h) that the directional band gap along $\overline{\Gamma}-\overline{M}$ direction always opens for $k_z < 0.4\pi/d$. Figures 4(h) and 4(i) are the corresponding projected bulk band and simulated transmission. We note that the transmission fringes in the passing band from 13 to 14 GHz are due to Fabry-Perot effect induced by the impedance mismatch between incident wave and the bulk mode. In our measured spectra of Fig. 4(j), the passing band from 10.5 to 11.8 GHz can be clearly seen. However, the upper passing band from 13 to 14 GHz cannot be excited due to the impedance mismatch. Together with the transmission spectra along $\overline{\Gamma}-\overline{K}$ direction, we found that a 2D complete band gap opens near the Weyl point frequency of 12.25 GHz in $k_x$-$k_y$ plane for nonzero $k_z$ and that the gap width broadens as $k_z$ increases. The calculated nonzero Chern number of this band gap indicates the existence of a chiral surface state, which will be confirmed in our surface measurement.

Owing to the topological charge of Weyl points, the surface states connecting the Weyl points with opposite charges are guaranteed by bulk-surface correspondence [35].

Consider a surface of the Weyl photonic crystal perpendicular to the $\overline{\Gamma}-\overline{M}$ direction (y-direction) bounded by a PEC slab, as shown in Fig. 5(a). Periodic boundary conditions are applied in both the x- and z-directions in our simulation. Surface dispersions near 12.25GHz are shown in Fig. 5(b). Surface states are plotted in color while the projected bulk states are plotted in gray. Only half of the Brillouin zone ($k_z \in [-0.5\pi/d, 0.5\pi/d]$) is shown for clarity. The Weyl points at K and K' with frequency of 12.25 GHz are projected onto $(k_x, k_z) = (2\pi/3a, 0)$ and $(4\pi/3a, 0)$, respectively. Two linearly-dispersive cones of projected bulk bands are formed near these points. Two sheets of surface states connected to the two Weyl points are found in positive $k_z$ region and negative $k_z$ region, where the colors indicate the frequency of surface state. The surface states with positive $k_z$ always have positive group velocity (i.e. in the +x direction), which is consistent with the $k_z$-Chern number in Fig. 1(j); while surface states with negative $k_z$ always have negative group velocity. When we see $k_z$ as an additional parameter of the 2D subsystem, Weyl points can be viewed as phase transition points between the $k_z$ planes with different Chern numbers. This can be seen by cutting three $k_z$ slices ($k_z = 0.05\pi/d, 0, -0.05\pi/d$) in the surface dispersion, as shown in Fig. 5(c-e). When $k_z = -0.05\pi/d$, the gap Chern number between the fifth and six bands is -1. The subsystem has an anticlockwise (in xy plane) surface state, as shown by the blue line in Fig. 5(e). When $k_z$ increases to 0 (Fig. 5(d)), the fifth and sixth bands touch at K and K'. The surface states are symmetric about $k_x$ due to time-reversal symmetry. When $k_z$ further increases to $0.05\pi/d$ (Fig. 5(c)), the 2D band gap reopens with Chern number of 1. The group velocity of surface states changed to clockwise. In addition, two other Weyl points at H and H' lie at $(2\pi/3a, \pi/d)$ and $(4\pi/3a, \pi/d)$ of the surface Brillouin zone, which are not shown in Fig. 5(b). Due to the band dispersion along the y-direction, these two Weyl points are immersed in the projected band of the bulk state in the surface Brillouin zone. The sheet of surface state with positive and negative kz will eventually merge into the bulk state (see the Supplemental Material).

One significant property of these chiral surface states are the robustness to the $k_z$-preserved scattering. For instance, we remove four PEC rods near the surface of Weyl photonic crystal (black circles in Fig. 6(a)). EM wave of 12.6 GHz is incident from the left with $k_z = 0.3\pi/d$. Since this line defect does not disrupt periodicity in z-direction, $k_z$ is preserved. The surface only supports a rightward surface mode for $k_z = 0.3\pi/d$. Thus EM wave can pass through this defect without backscattering, as shown by the $E_z$ field pattern in Fig. 6(a). Figure 6(b) gives another example with a PEC bar inserted into the bulk of photonic crystal. EM wave can wrap round this defect and keep moving rightward. In both simulations, photonic crystals with infinite height are considered as periodic boundary condition is used in z-direction. To test the robustness of the surface state, samples with these defects were fabricated. Figure 6(c) and 6(d) show the photographs of samples, both have 60 periods (about 32.4cm) in z-direction and are capped by a 1mm-thicked aluminum slab. EM wave was impinged onto the left of the sample with incident angle $\theta$. The incident beam width in z-direction is about 22 cm. Figures 6(e) shows the results for the sample with missing rods when incident angle is 35° ($k_z = 0.2581\pi/d$ for 12.5 GHz). Another horn antenna with the same tilted angle was put on the right to receive the transmitted wave. We measured transmittances for the cases with 1, 2 and 4 missing rods (red, blue and green curves in Fig. 6(e)). Figure 6(c) shows the configuration of 4 missing rods. For the case of 2 missing rods, only the two rods nearest to the surface were removed. Compared to the transmittance of the sample without defect (black curve), we found that the spectra overlapped with each other in the frequency range from 12.2 to 12.65 GHz (light cyan box). It indicates that no backscattering was introduced by the removed aluminum rods in the nontrivial band gap with nonzero $k_z$. However the bandwidth of robust transport is narrower than the band gap (from 11.8 to 12.8 GHz) predicted by bulk band structure. This is due to the finite size of beam width of incident wave packet, which leads to the spread of the Fourier components around $k_z = 2\pi f \sin\theta / c$. Purple curve in Fig. 6(f) plots the measured transmittance when an aluminum bar was inserted into the crystal. Robust transport was also observed from 12.2 to 12.65 GHz. From Figs. 1(g-i), the gap width of this nontrivial band gap increases as $k_z$ increases when $k_z < 0.4\pi/d$. Figures 6(g) and 6(h) show the measured transmittances for the incident angle of $\theta = 50°$ ($k_z = 0.3447\pi/d$ for 12.5 GHz). Similar robust transmissions were observed from 12.3 to 13 GHz (highlighted by the light cyan box), which is broader than the case of $\theta = 35°$. Nevertheless this gap width is narrower than the expected range (from 11.8 to 13 GHz) due to the finite beam width.

In conclusion, we designed and fabricated a microwave Weyl photonic crystal which carries both single and double Weyl points.. Weyl point and nontrivial bulk band gap for a fixed $k_z$ were confirmed by numerical simulations and measured angle-resolved transmissions. Topologically-protected chiral surface state between Weyl photonic crystal and PEC were also measured and demonstrated to be robust against $k_z$-preserved scatterers.

This work was supported by Hong Kong Research Grants Council (grant no. AoE/P-02/12).

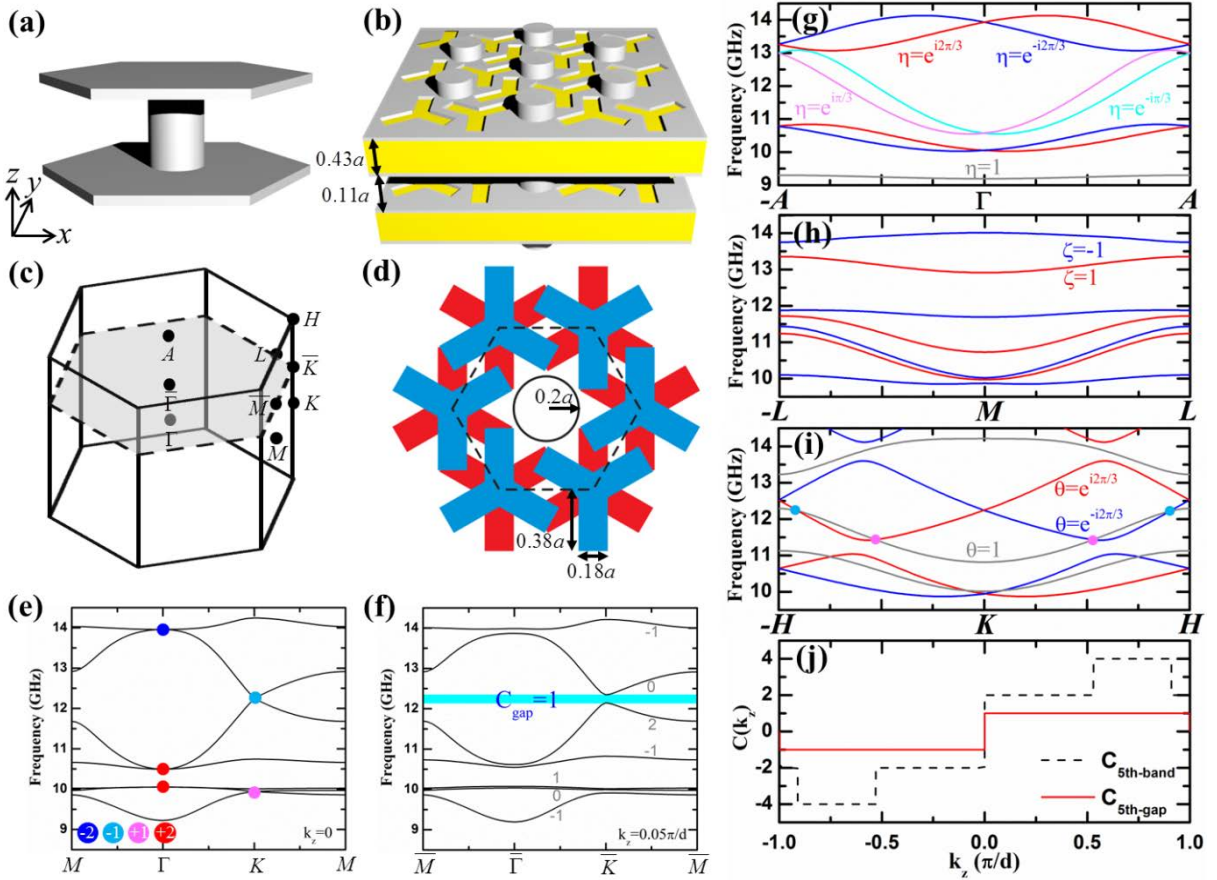

**FIG. 1. (color online) Realization of Weyl points in EM system by introducing interlayer coupling. (a) The unit cell of a single layer system built from a hexagonal array of PEC cylinders bounded by two PEC slabs, which has a conical dispersion at the 2D Brillouin zone corner. This can be realized with metal coated PCB board that is pierced by a hexagonal array of aluminum rods. PCB boards stacked in z-direction forms a 3D photonic crystal. (b) Multilayer system built from PCB boards stacked in z-direction. Interlayer couplings are introduced by the Y-shaped slots on both sides of the PCB boards. (c) Reciprocal space of a hexagonal lattice. (d) Top view of the unit cell (dashed hexagon) of the multilayer system. Blue and red areas highlight the Y-shaped slots on the upper surface and the lower surface of the PCB board. Since the photonic crystal has translational symmetry along z-direction, $k_z$ is a good quantum number. The system**

with a fixed $k_z$ has a 2D band structure in the reduced Brillouin zone (gray plane in (c)). Chern numbers are well defined on each $k_z$ slice. Weyl point can viewed as the phase transition point between the $k_z$ slices with different Chern numbers. (e) Bulk band structure on the $k_z$=0 plane. It has several Weyl points with different charges (highlighted by different colors). (f) Bulk band structure on the $k_z$=0.05$\pi$/$d$ plane. (g)-(i) Dispersion along the z-direction at $\bar{\Gamma}$, $\bar{M}$ and $\bar{K}$, respectively. Bands with different rotation eigen value are plotted in different colors. Since the change of rotation eigenvalue results in the change of band (gap) Chern number, each crossing point in (g) or (i) is a Weyl point, whose charge depends on the ratio between two band's rotation eigen values. Four Weyl points between the 4th and 5th bands, which induce the jumps of the 5th band Chern number in (j), are highlighted in (i). (j) Chern numbers of the 5th gap (red solid line) and the 5th band (black dashed line) as a function of $k_z$, the jump of which implies the topological charge of associated Weyl points.

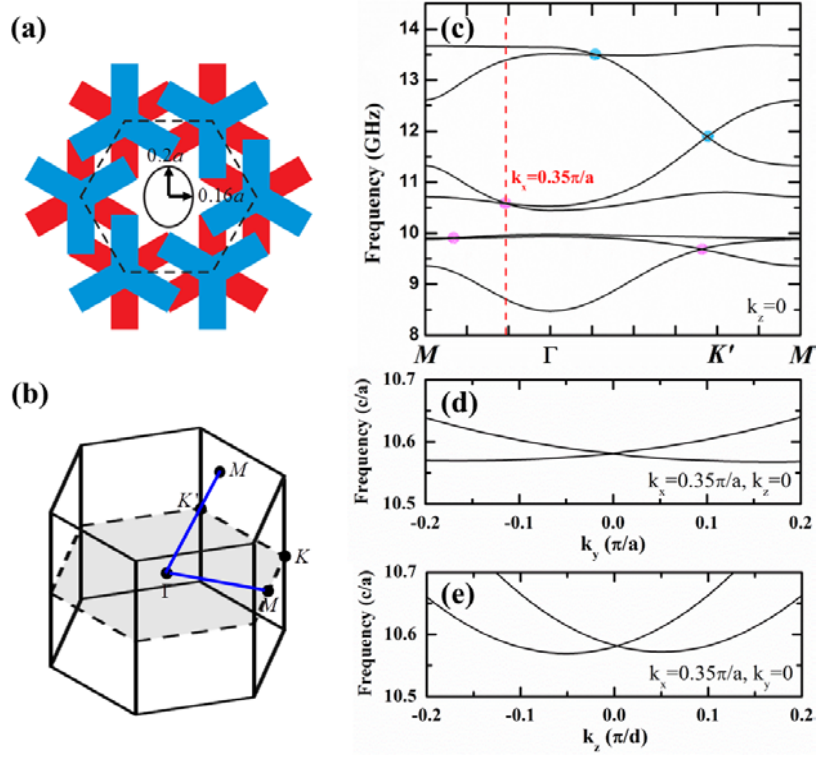

FIG. 2. (color online) Double Weyl points breaking into single Weyl points due to symmetry breaking. (a) Unit cell of the C$_3$ symmetry broken crystal where the PEC cylinder in Fig. 1(d) is replaced by an elliptical cylinder. (b) Corresponding Brillouin zone. Blue solid line highlights the path of k-point when calculating the dispersion in (c). The upper M point, which is related to the lower M point by a reciprocal lattice vector, lies on the Γ−K′ direction. (c) Band structure on the plane of $k_z = 0$. Each of the three double Weyl points in Fig. 1(e) splits into two single Weyl points. Only three of them are shown (the band crossing points between second and third, fourth and fifth, sixth and seventh bands). The other three single Weyl points can be inferred by applying C$_2$ rotation about the axis of elliptical cylinder. (d) & (e) show the linear dispersions near the Weyl point between fourth and fifth bands along the y- and z-directions, respectively.

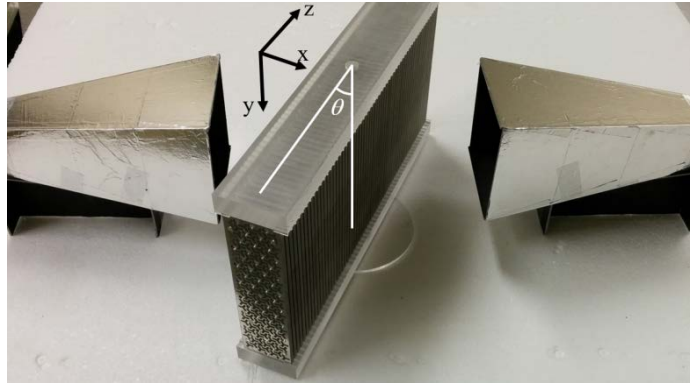

**FIG. 3. (color online) Experimental setup for bulk transmission measurements. EM waves are emitted and received by two horn antennas with electric field lying in xz plane. By changing the incident angle $\theta$, bulk states with different $k_z=2\pi f \sin\theta/c$ can be excited.**

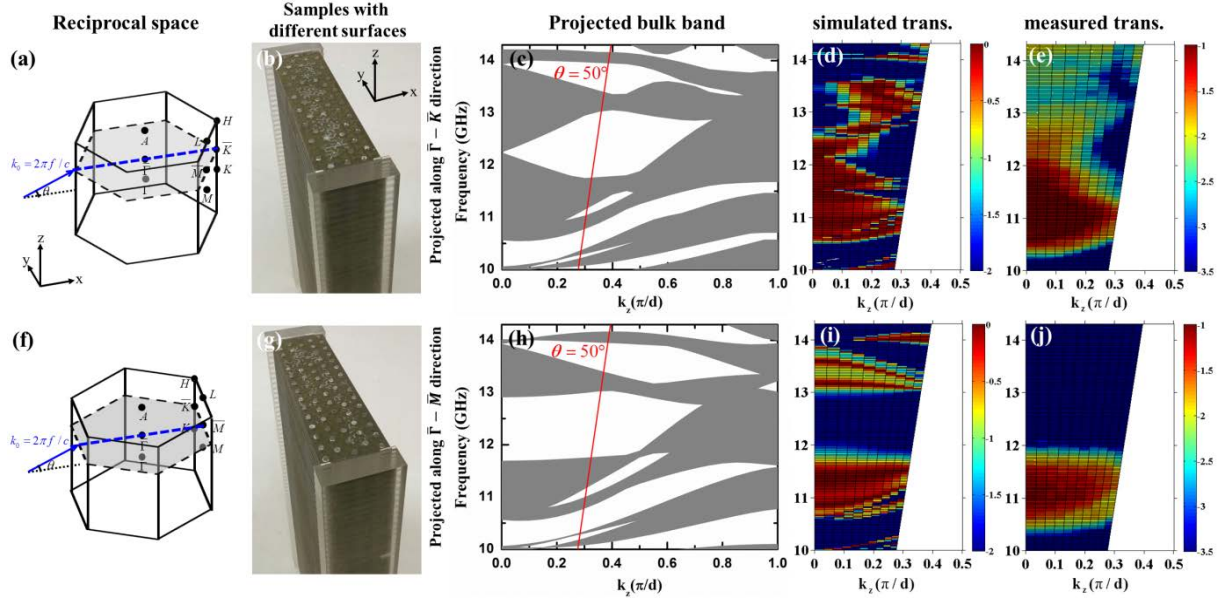

**FIG. 4. (color online) Projected band structure and bulk transmission. (a) Reciprocal space corresponding to the sample shown in (b). (b) Photograph of the sample with surface perpendicular to the Γ−K direction. Blue arrow in (a) indicates the incident wave vector in the *xoz* plane with tilted angle *θ*. The bulk modes lying at the blue dashed line, which is parallel to the $\bar{\Gamma}-\bar{K}$ direction, can be excited. (c) Bulk band projected along the $\bar{\Gamma}-\bar{K}$ direction as a function of $k_z$. The red line depicts the $k_z$ corresponding to the maximal incident angle ($\theta = 50°$) in our measurement. (d)/(e) simulated and measured transmission spectra along the $\bar{\Gamma}-\bar{K}$ direction. (f) Reciprocal space corresponding to the sample shown in (g). (g) Sample with surface perpendicular to the Γ−M direction. (h) Bulk band projected along the $\bar{\Gamma}-\bar{M}$ direction. (i)/(j) simulated and measured transmission spectra along the $\bar{\Gamma}-\bar{M}$ direction.**

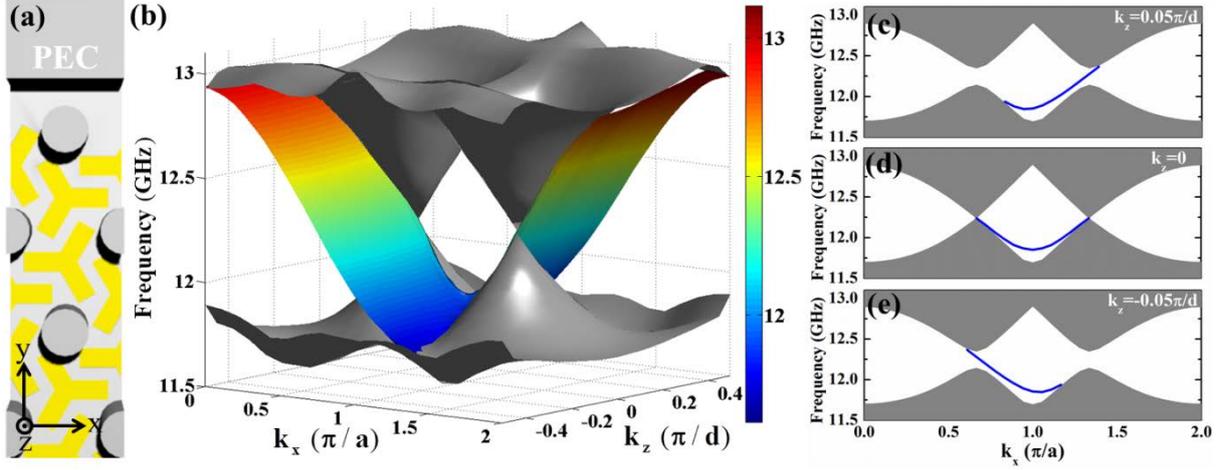

**FIG. 5. (color online) Surface dispersion between Weyl photonic crystal and PEC. (a) Configuration of the Weyl photonic crystal bounded by a PEC slab. (b) Calculated surface dispersion in the surface Brillouin zone. For clarity, half of the Brillouin zone ($k_z \in [−0.5\ π/d, 0.5\ π/d]$) are plotted. Projected bulk states are plotted in gray while the surface states are plotted in color. Two linear cones of bulk states lying at $k_z$=0 plane are due to the two Weyl points at K and K' (where the 5th and 6th bands intersect) with topological charge of -1. The two Weyl points at H and H' with charge of +1, which lie at $k_z$=π plane, are not shown. (c)-(e) Surface dispersions at $kz$=0.05π/d, $kz$=0 and $kz$=−0.05π/d, respectively, where the blue lines denote the surface states.**

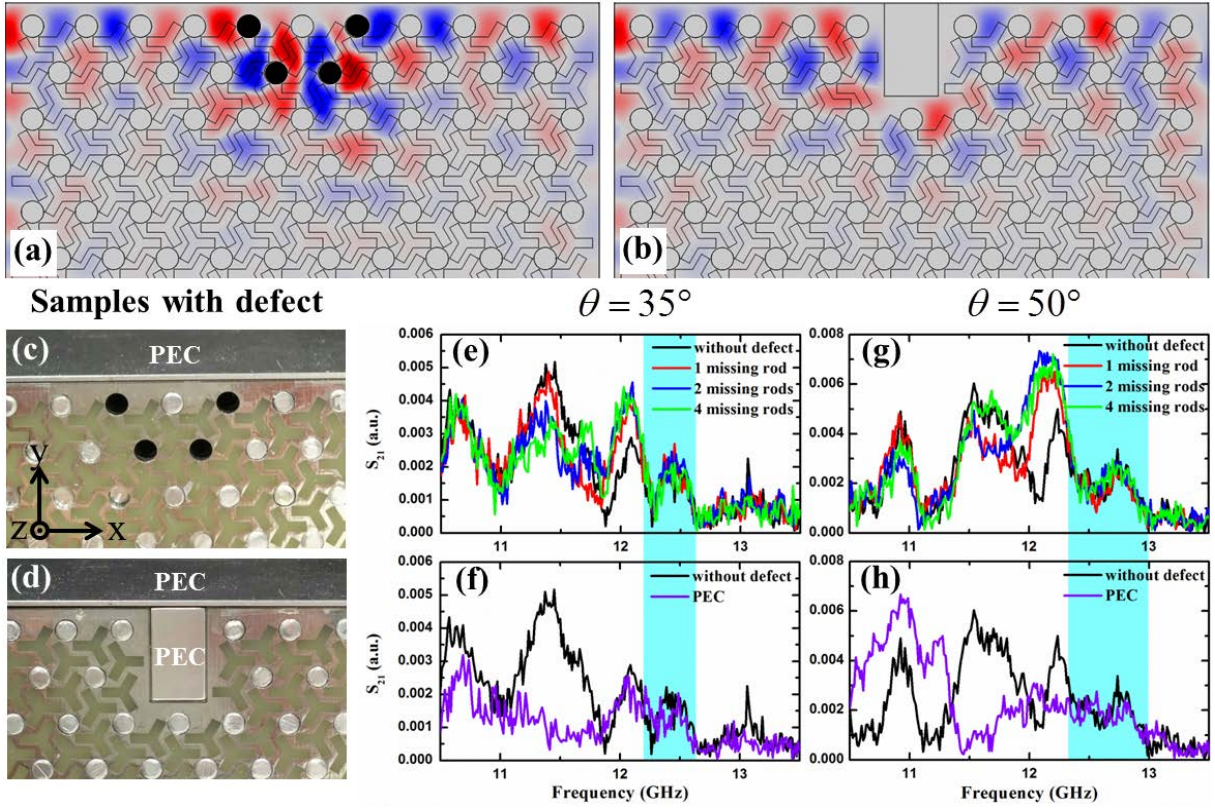

FIG. 6. (color online) Robust surface states between Weyl photonic crystal and PEC. For a fixed nonzero $k_z$, the surface only supports surface states in one direction. Hence the surface states are robust against scattering that preserves $k_z$. (a) Simulated $E_z$ field pattern when 4 PEC rods (highlighted by black circles) near the surface are removed. (b) Simulated field pattern in the presence of a PEC bar. In both cases, EM waves with $k_z=0.3\pi/d$ are incident from the left. Since both kinds of defects in (a) and (b) are periodic in z-direction and $k_z$ is conserved, EM wave can pass through or wrap around the defect without backscattering. (c)/(d) Photograph of the samples with defect. (e) and (f) show the measured transmission spectra for two kinds of defect when the incident angle $\theta=35°$ ($k_z=0.2581\pi/d$ for 12.5GHz). (g) and (h) show the results for $\theta=50°$ ($k_z=0.3447\pi/d$ for 12.5GHz). Cyan boxes in (e)-(h) highlight the frequency region of $k_z$-preserved one-way surface states where no obvious backscattering are introduced by the two kinds of defects. Note that the frequency region of robust transport for $\theta=50°$ is wider than the case of $\theta=35°$. This is consistent with the fact that the nontrivial bandgap in reduced 2D Brillouin zone is wider for lager $k_z$.